\documentclass[runningheads, 12pt]{mmef} 
\usepackage[a4paper,top=3.8cm,bottom=3.8cm,left=2.5cm,right=2.5cm]{geometry} 
\usepackage[]{latexsym}\usepackage{amssymb}\usepackage{amsmath}\usepackage{lscape}\usepackage[english]{babel} \usepackage{fancyhdr}\usepackage{appendix}\usepackage{graphicx}\usepackage{amsfonts, wasysym}\input{psfig.sty}
\usepackage{setspace}
\begin{document}
\setcounter{page}{1}
\title{Endogenous Bubbles in Derivatives Markets:\\The Risk Neutral Valuation Paradox} 
\titlerunning{Endogenous Bubbles in Derivatives Markets\dots}

\author{Alessandro Fiori Maccioni\thanks{I wish to thank Prof. Marco Vannini for being my mentor and Prof. Angelo Antoci for considerate 
guidance during my studies. I also wish to thank Prof. Martin Schweizer for precious comments on a previous draft and 
Dr Andrea Pinna and Dr Claudio Detotto for useful talks on the seminal stages of the 
research. This work is part of a joint research program sponsored by the Department of Economics of the University of Sassari and the Regione Autonoma della Sardegna under the regional law n.  7/2007. I also gratefully acknowledge the support to research provided by the Doctoral School in Law and Economics of the University of Sassari.}}

\institute{CRENoS - Centre for North-South Economic Research\\
and Dipartimento di Economia, Impresa e Regolamentazione,\\Universit\`a di Sassari,\\Via Torre Tonda n. 34 - 07100 Sassari, Italy\\\email{alex.fiori.maccioni@uniss.it}
}

\maketitle\begin{abstract}
This paper highlights the role of risk neutral investors in generating endogenous bubbles in derivatives markets.
We find that a market for derivatives, which has all the features of a perfect market except completeness and has some risk neutral investors, can exhibit extreme price movements which represent a violation to the Gaussian random walk hypothesis.
This can be viewed as a paradox because it contradicts wide-held conjectures about prices in informationally efficient markets with rational investors.
Our findings imply that prices are not always good approximations of the fundamental values of derivatives, and that extreme price movements like price peaks or crashes may have endogenous origin and happen with a higher-than-normal frequency.

\noindent {\bf Keywords.}
Risk neutral, martingale, derivatives, efficient market, bubble.

\noindent {\bf M.S.C. 2010 classification.} \textsc{\small 60G, 60H, 91B, 91G.}

\noindent {\bf J.E.L. classification.} \textsc{\small \upshape{Primary:} C73, G12, G13.
\upshape{Secondary:} C78, G01.}
\end{abstract}

\section{Introduction}
The present study addresses the role of risk neutral investors in generating turmoil in financial markets. Our hypothesis is that, in derivatives markets, random trades made by risk neutral investors and based on arbitrage free valuation can move prices away from their fundamental values. This effect can improve market instability by leading to abrupt price adjustments towards their fundamental values. These adjustments can take the form of either price peaks or price crashes, have an endogenous origin and can happen systematically.

Our main result consists in the demonstration of the existence of a price anomaly, that we have called the risk neutral valuation paradox.
The paradox is a consequence of the fundamental theorem of asset pricing and the related martingale theory of bubbles. Following these theories, prices in a derivatives market can diverge from their fundamental values if and only if the market is incomplete. Thus, we intend to demonstrate our assumption in a market that has all the properties of a perfect market except completeness.



We define a competitive economy of pure exchange with uncertainty. We consider a single consumption good, which acts as num\'eraire. Agents are interested in certain consumption at present time and in state contingent consumption on future dates. There is an underlying asset, composed by the sequence of realizations $x_{t}(\omega)$ of the stochastic process $X_t$. A realization $x_{t}(\omega)$ represents the contingent consumption gained or lost at date $t$ by the owner of the underlying asset if the state is $\omega$. Tradable assets consist in the underlying asset and the set of its derivatives. We consider continuous trading dates. The market is incomplete so that price bubbles might come into existence.

Agents aim only at maximizing their expected utility. They differ for risk attitudes and can be divided into the two groups of risk averse and risk neutral. 

Risk averse agents will determine, for each tradable consumption bundle and according to their subjective expected utility, 
a reservation price which will constitute a bound for their willingness to trade. 
We name their strategy as `fundamental trading'. 

Risk neutral agents, under some assumptions, are indifferent between the choices of buying, selling or not trading at all any consumption 
bundle, and will switch from one action to the other randomly over time. However, given their choice of buying or selling, 
they will be available to, respectively, bid more or ask less than the current market price, because the choice affects 
their price expectation through Bayesian updating. We name their strategy as `technical trading'.



We show that, in a market where all agents operate as fundamental traders, the price process is \textit{bounded}. We define such market as `non speculative market'.

Once that an equilibrium price is reached, 
risk neutral agents can start to operate as technical traders.
We demonstrate that the entrance of technical traders in the market leads the price process to follow a \textit{locally unbounded} Brownian motion; thus, there exists a positive probability that the market price may exit from the boundaries that previously existed when only fundamental traders were operating.

When the market price stays inside the boundaries of a non speculative market, we define the market conditions as `normal'.
When it goes outside these boundaries, we define the market conditions as either `depression' or `bubble', depending if the price stays, respectively, below or above the boundaries. In these cases, the market bears the risk of an abrupt price adjustment towards its fundamental values.
Indeed, if risk neutral agents stop to operate as technical traders, even for an instant, the market becomes again non speculative and the price goes back into the boundaries via an abrupt price adjustment. 
These adjustments can take the form of either price peaks or crashes and their frequency will sum up to the normal price movements, thus affecting the tails of the distribution of price movements.

The paper starts with a brief exposition of the theoretical foundation (sect.\,\ref{foundation}) and the mathematical formalization of the economic environment (sect.\,\ref{environment}). 
The paper follows with the definition of the model and the demonstration of the paradox (sect.\,\ref{model}). 
Finally, conclusions are drawn (sect.\,\ref{conclusions}).

\section{Theoretical Foundation} \label{foundation}
The paradox that we propose is a consequence of what is known as the `fundamental theorem of asset pricing', which is at the core of the arbitrage pricing theory.
The fundamental theorem states that the absence of free lunch in a frictionless market is equivalent to the existence of an equivalent martingale probability measure, under which all assets have the same expected return, and that this measure is unique if and only if the market is complete.


The foundations of the arbitrage pricing theory in modern 
 finance are provided by Black and Scholes \cite{bla:sch} and by Merton \cite{mer} in their celebrated contributions on option valuation.
Cox and Ross \cite{cox:ros} introduce the concept of risk neutral valuation and argue that in a market with no arbitrage opportunities, it is possible to reassign probabilities to give all assets the same expected returns.
Harrison and Kreps \cite{har:kre}, Kreps \cite{kre} and Harrison and Pliska \cite{har:pli} make a breakthrough in arbitrage pricing theory and give a rigorous foundation to the fundamental theorem of asset pricing. They use It\^o calculus to define the concept of no-arbitrage,
refer to risk neutral probabilities as the `equivalent martingale measure', and demonstrate the fundamental theorem for trading strategies that are simple integrands.
These results were extended in various directions, among the others\footnote{
See also: Dybvig and Huang \cite{dyb:hua}, Duffie and Huang \cite{duf:hua}, Stricker \cite{str}, Ansel and Stricker \cite{ans:str} and Lakner \cite{lak}.}, 
by Dalang, Morton and Willinger \cite{dal:mor:wil}, while Delbaen \cite{del} and Schachermayer \cite{sch93} prove the theorem for special cases.\footnote{
Simple proofs are also proposed by Schachermayer \cite{sch92}, Kusuoka \cite{kus}, Kabanov and Kramkov \cite{kab:kra} and Rogers \cite{rog}.}

A general version of the fundamental theorem of asset pricing is provided by Delbaen and Schachermayer \cite{del:sch94} \cite{del:sch98}. They extend the theorem for trading strategies that are general integrands. Furthermore, they decline the no-arbitrage condition under the fundamental concept of `no free lunch with vanishing risk' and they show that such condition is fulfilled if and only if there exists an equivalent probability measure under which the price process is a sigma martingale or, in the continuous case, a local martingale.

The literature and theories discussed above are at the base of what is sometimes called the `martingale theory of bubbles'.\footnote{See for example Jarrow and Protter \cite{jar:pro}.}
Harrison and Kreps \cite{har:kre78} provide the classical definition of the fundamental value of an asset as the value of its discounted cash flows under the equivalent martingale measure. Therefore, a bubble would originate from the deviation between the fundamental value of an asset and its market price.
This theory focuses on the characteristics of asset bubbles and the evaluation of derivative securities in economies which satisfy the no arbitrage condition.

The theory of bubbles in the classical economic framework requires considerably stricter conditions than in the martingale approach. Indeed, in economic equilibrium models, the price function requires a precise definition of additional characteristics of the economy, such as the functions of supply and demand.\footnote{
We recall the contributions on asset bubbles in classical economic frameworks proposed by Tirole \cite{tir82} and Santos and Woodford \cite{san:woo} on markets with finite trading horizon and rational expectations, by Tirole \cite{tir85}, O'Connell and Zeldes \cite{oco:zel} and Weil \cite{wei} on markets with rational traders and infinite trading horizon, and by De Long et al. \cite{del:shl:sum:wal} on markets where there are irrational traders.  For good reviews, see Camerer \cite{cam}, Scheinkman and Xiong \cite{sch:xio}, and Jarrow, Protter and Shimbo \cite{jar:pro:shi10}.}

In the martingale theory of bubbles, the no arbitrage condition is often imposed in the form of `no free lunch with vanishing risk' (NFLVR) proposed by Delbaen and Schachermayer \cite{del:sch94}.
In the models proposed by Loewenstein and Willard \cite{loe:wil1} \cite{loe:wil2}, by Cox and Hobson \cite{cox:hob}, and by Heston, Loewenstein and Willard \cite{hes:loe:wil}, bubbles can violate numerous classical option pricing theorems including the put-call parity. This result is only partially supported by empirical evidence.\footnote{
See for example Kamara and Miller \cite{kam:mil}, and Ofek and Richardson \cite{ofe:ric}.}
To overcome this limitation, Jarrow, Protter and Shimbo \cite{jar:pro:shi06} \cite{jar:pro:shi10} impose the further condition of no dominance\footnote{
See Merton \cite{mer}.} of portfolios. This condition is stronger that the NFLVR but still substantially weaker than imposing a market equilibrium.
 They show that the addition of the no dominance condition excludes all asset price bubbles in complete markets with infinite trading horizons, because the fundamental values of assets and their market prices should always be identical. Consequently, if bubbles are to exist, the market should be incomplete.

According to the fundamental theorem of asset pricing, an incomplete market presents a wide range of different martingale measures that may be used for estimating the fundamental values of assets. Eberlein and Jacod \cite{ebe:jac} suggest that these martingale measures and, consequently, the prices of derivatives should have a closed range of variation. Schweizer and Wissel \cite{sch:wis}, and Jacod and Protter \cite{jac:pro} argue that market prices of derivatives can reveal which one of these martingale measures is the one currently adopted by the market. Then, a bubble can start (from a situation of non-bubble) when traders decide to adopt a different martingale measure. In economic terms, this variation will correspond to a regime change in the core values of the economy (endowments, beliefs, risk aversion, technologies, institutional structures). In financial terms, the change in the martingale measure can leave unchanged the price of the underlying security but, as the market is incomplete, it will change the price of some of its derivative securities. 
To investigate the birth of a bubble or its presence in the market is therefore essential to analyze temporal trends in the prices of derivative securities,\footnote{See also Jarrow, Kchia and Protter \cite{jar:kch:pro}.} as we do in this paper.

The contribute of this paper to the theory of financial markets is to highlight the role of risk neutral investors in generating endogenous bubbles in derivatives markets. 
In the following sections, we propose and demonstrate an original paradox of asset pricing. The paradox suggests that almost perfect derivatives markets may exhibit extreme price movements, 
which represent a violation to the Gaussian random walk hypothesis, have an endogenous origin and may happen with higher-than-normal frequency.


\section{The Economic Environment} \label{environment}
\subsection{Basic Definitions} \label{basic:definitions}
Let us consider a competitive economy of pure exchange with uncertainty.
We are given a continuous set of trading dates $t\in[\,0,\;T\,]$ where $t$ are the instants of time at which market participants can trade.
There is a single consumption good, which acts as num\'eraire.

Agents are interested in certain consumption at present time and in state contingent consumption on future dates.
Future consumption is represented by functions of the underlying asset $\tilde{\mathbf{x}}_{t}$, which is the vector composed by the sequence of realizations $x_{t}(\omega)$ of the stochastic process $X_t$. The realization $x_{t}(\omega)$ represents the contingent consumption gained or lost at date $t$ by the owner of the underlying asset $\tilde{\mathbf{x}}_{t}$ if the state is $\omega$.
The underlying asset and its possible derivative functions form the convex set $\mathbf{x}_{t}^{set}$, which represents the family of tradable vectors $\mathbf{x}_t$ that can be created and exchanged at date $t$. In other words, $\mathbf{x}_{t}^{set}$ represents the securities traded in the market at any given date $t$. To improve readability, we will write $\mathbf{x}_{t}$ for the generic element of $\mathbf{x}_{t}^{set}$.

At each trading date $t$, agents can exchange any vector of future contingent consumption $\mathbf{x}_{t}\in \mathbf{x}_{t}^{set}$ with units of certain present consumption $r_t$.
Thus, we consider consumption bundles of the form $(r_t,\mathbf{x}_{t})$ where the real number $r_t$ represents the units of certain consumption at the present date $t$ and the vector $\mathbf{x}_{t}$ represents the units of state contingent consumptions over the interval $(\,t,\,T\,]$.

\subsection{Probability Assumptions} \label{probability:assumptions}
Let $(X_t)_{t\in[0,T]}$ be an adapted c\`adl\`ag stochastic process on the probability space $\left( \,\Omega,\, \mathbb{F},\, \mathcal{P}\,\right)$, where:
\begin{itemize}
	\item the universal sample space $\Omega$ is the set of all possible elementary outcomes $\omega \in \Omega$;
	\item the filtration $\mathbb{F}$ is the set of the $\sigma$-algebras 
	 $\mathbb{F}=\left\{\mathcal{F}_t\right\}_{t \in [0,T]}$, where $\mathcal{F}_t$ represents the information available at any time $t$;
	\item $\mathcal{P}$ is the probability measure on the universal sample space $\Omega$.
\end{itemize}

The underlying probability space that we define fulfills the usual hypotheses of the general theory of financial asset pricing.
We can interpret $\mathcal{F}_t$ as the questions that agents can answer at time $t$ regarding past and present states of the world. The information becomes more and more precise (i.e. the set of measurable events increase) as new events from the present becomes known. 

For the purpose of the model, we assume that the information is continuously and completely available to each individual (and, accordingly, that the filtration is right continuous).
Furthermore, we assume that information is free. Thus, no additional costly information is available and the model avoids any potential application of the Grossman and Stiglitz paradox \cite{gro:sti} on the impossibility of informationally efficient markets.
We also assume, with no loss of relevance, that investors have common prior beliefs on the probability of future events.

\subsection{Market Participants} \label{agents}
We define rationality in strictly axiomatic form, according to von Neumann and Morgenstern \cite{von:mor} and Savage \cite{sav}. Furthermore, we assume that all the individuals operating in the market are rational.

We denote the set of conceivable agents with $\mathbf{A}$. The $i$-th agent is characterized at each date $t$ by an initial endowment $(\hat{r}_{t}^{i},\hat{\mathbf{x}}_{t}^{i})$ and by a preference relation~\,${\succsim}_i$\, over the space of consumption bundles $(r_t,\mathbf{x}_t)$. Preferences depend on the subjective degree of risk aversion and are assumed to be complete, transitive, convex, continuous, strictly increasing and independent. 
Following these properties, the agent is able to express the preferences towards consumption bundles via an additive von Neumann-Morgenstern utility function $u_i:(r_t,\mathbf{x}_t)\rightarrow\mathbb{R}$.

Agents aim only at maximizing their expected utility and choose the strategies that they repute best performing.
There are two possible alternative strategies: `fundamental trading' and `technical trading', both consisting of simple integrands.


Fundamental traders consist of risk averse agents. For each tradable consumption bundle, they estimate a reservation price according 
to their subjective expected utility. Then, they will be available to make any trade that is available in the market 
at more favourable conditions than its reservation price.

Technical traders consist of risk neutral agents who assume that other risk neutral agents are operating and that the price process follows a Brownian motion. Under these assumptions, they will be indifferent between the choices of buying, selling or not trading at all a given consumption bundle, and will switch from one action to the other randomly over time.
However, given their choice of either buying or selling, they will be available to, respectively, bid more or ask less than the current market price, because the choice affects their price expectation of the consumption bundle through Bayesian updating.



Although each agent can both buy and sell, we prefer to consider each trader as two different individuals, 
one that exclusively buy and the other that exclusively sell.
Then, from the intersection of the different categories, we can formally define the set of conceivable agents 
$\mathbf{A}:\lbrace \mathbf{FB},\,\mathbf{FS},\,\mathbf{TB},\,\mathbf{TS} \rbrace$ as composed by the distinct subgroups of 
fundamental traders that are buyers, $\mathbf{FB}$, and sellers, $\mathbf{FS}$, and by the technical traders that are buyers, 
$\mathbf{TB}$, and sellers, $\mathbf{TS}$.

As said above, the number of technical traders varies randomly over time depending on the choices of risk neutral investors between 
indifferent alternatives  (buying, selling, not trading). 
Thus, we will also use, when convenient, the notation $\mathbf{TB}_t$ and $\mathbf{TS}_t$ to denote the sets of risk neutral agents 
operating as technical traders (respectively, buyers and sellers) at a specific time $t$.

\subsection{Trading System} \label{trading}
Exchanges happen within an electronic trading system. Agents submit orders to the electronic system, which immediately searches for matching orders; in case they exist, the system executes the trade and remits the num\'eraire to the traders. To be more precise, when a consumption bundle $(r_t,\mathbf{x}_{t})$ is traded, the system remits immediately the payment $r_t$ to the seller and, from then inwards, it pays or subtracts to the buyer (as it accrues) the contingent consumption generated by $\mathbf{x}_{t}$. Agents go bankrupt and must exit from the market if their wealth represented by units of certain consumption is wiped out by losses.

\label{sec:orders}
Agents' trading orders can be of four types:
\begin{itemize}
\item the limited order of buying, $\text{\textit{Bid}}_{i}(\mathbf{x}_{t}\!)$, which expresses the \textit{maximum} amount of certain present consumption that an agent $i\in\mathbf{FB}$ is willing to offer for buying a vector of future contingent consumption $\mathbf{x}_{t}$;
\item the limited order of selling, $\text{\textit{Ask}}_{i}(\mathbf{x}_{t})$, which expresses the \textit{minimum} amount of certain present consumption that an agent $i\in\mathbf{FS}$ is willing to accept for selling a vector of future contingent consumption $\mathbf{x}_{t}$;
\item the market order of buying, $\text{\textit{Buy}}_{i}(\mathbf{x}_{t})$, which expresses the will of an agent $i\in\mathbf{TB}$ to buy a given vector of future contingent consumption $\mathbf{x}_{t}$ at \textit{any} price, as long as it is the market-clearing price $\text{\textit{Equil}}\,(\mathbf{x}_{t})$ plus a small sum~$\epsilon$;
\item the market order of selling, $\text{\textit{Sell}}_{i}(\mathbf{x}_{t})$, which expresses the will of an agent $i\in\mathbf{TS}$ to sell a given vector of future contingent consumption $\mathbf{x}_{t}$ at \textit{any} price, as long as it is the market-clearing price $\text{\textit{Equil}}\,(\mathbf{x}_{t})$ minus a small sum~$\epsilon$.
\end{itemize}

We assume that the market orders of buying and selling, $\text{\textit{Buy}}(\mathbf{x}_{t})$ and $\text{\textit{Sell}}(\mathbf{x}_{t})$, put, respectively, upward and downward pressure on the market-clearing price, $\text{\textit{Equil}}\,(\mathbf{x}_{t})$. The final effect on the price at time $t+dt$ depends on which of the two sets of technical traders, buyers $\mathbf{TB}$ or sellers $\mathbf{TS}$, is larger at time $t$.

The element $\epsilon$ can be interpreted as the Bayesian updating of price expectations of technical agents due to their choice of buying or selling. We can also interpret $\epsilon$ as a small sum that agents should add or subtract to the market-clearing price $\text{\textit{Equil}}\,(\mathbf{x}_{t})$ to make sure that their market orders of buying or selling, $\text{\textit{Buy}}(\mathbf{x}_{t})$ and $\text{\textit{Sell}}(\mathbf{x}_{t})$, would be executed.

The market-clearing price, $\text{\textit{Equil}}\,(\mathbf{x}_{t})$ 
 is determined according to a price function of the form:
\begin{equation} \label{complete:price:function}
\text{\textit{Equil}}(\!\mathbf{x}_t\!):= f \,\Bigl\{ \text{\textit{Bid}}_{\mathrm{FB}}(\!\mathbf{x}_t\!)\hspace{-1pt},\hspace{1pt} \text{\textit{Ask}}_{\mathrm{FS}}(\!\mathbf{x}_t\!)\hspace{-1pt},\hspace{1pt} \text{\textit{Buy}}_{\mathrm{TB}}(\!\mathbf{x}_t\!)\hspace{-1pt},\hspace{1pt} \text{\textit{Sell}}_{\mathrm{TS}}(\!\mathbf{x}_t\!)\hspace{-1pt},\hspace{1pt} \text{\textit{Equil}}(\!\mathbf{x}_{t-dt}\!) \Bigr\} \,\mapsto\, \mathbb{R},
\end{equation}
\noindent where $\text{\textit{Bid}}_{\mathrm{FB}}(\mathbf{x}_t)$ and $\text{\textit{Ask}}_{\mathrm{FS}}(\mathbf{x}_t)$ are the sets of trading orders from fundamental traders that are, respectively, buyers and sellers, and $\text{\textit{Buy}}_{\mathrm{TB}}(\mathbf{x}_t)$ and $\text{\textit{Sell}}_{\mathrm{TS}}(\mathbf{x}_t)$ are the sets of trading orders from technical traders that are, respectively, buyers and sellers.

The price function in formula (\ref{complete:price:function}) is expressed in general form only for the purpose of demonstrating 
the existence of the paradox. 
A more precise definition will be necessary to extend the results and define carefully how the random variations of the prices look over time.

\section{The Model} \label{model}
In the present section, we intend to demonstrate that, when no risk neutral investor is operating as technical trader, 
the price function in formula (\ref{complete:price:function}) becomes bounded. On the other hand, when some risk neutral investors are operating in the market as technical traders, the price function follows a locally unbounded Brownian motion.

Then, the demonstration of the risk neutral valuation paradox comes from the alternation of boundedness and unboundedness of the price process, due to risk neutral investors who switch randomly over time from fundamental to technical trading and vice versa.

\subsection{Basic definitions}
\begin{definition}[expected utility] \label{utility} The expected utility of a given consumption bundle $(r_t,\,\mathbf{x}_{t})$ for the $i$-th agent is the function:
\begin{equation} \label{equivalence}
U_i(r_t,\mathbf{x}_{t}) = E\,\big[\,u_i(r_t,\mathbf{x}_{t})\,\big] = \int_{\Omega}{u_i\,\big(\,r_t,\,\mathbf{x}_{t}(\omega)\,\big)\,\mathcal{P}\,d{\omega}}, \qquad \forall \,i \in \mathbf{A},
\end{equation}
\noindent where $u_i\big(r_t,\, \mathbf{x}_{t}(\omega)\big)$ is the utility of the $i$-th agent conditional on the realization of the event $\omega\hspace{-3pt}\in\hspace{-3pt}\Omega$, and $\mathcal{P}$ is the probability measure on the space of events $\Omega$.
\end{definition}

\begin{lemma}[preference between consumption bundles] \label{preference}
Let $(\,\dot{r}_{t},\,\dot{\mathbf{x}}_{t}\,)$ and $(\,\ddot{r}_{t},\,\ddot{\mathbf{x}}_{t}\,)$ be two different consumption bundles. The $i$-th agent strictly prefers the first bundle to the second if and only if:
\begin{equation}
(\,\dot{r}_{t},\,\dot{\mathbf{x}}_{t}\,) \; {\succ}_i \; (\,\ddot{r}_{t},\,\ddot{\mathbf{x}}_{t}\,)
\quad \Longleftrightarrow \quad U_i\,(\,\dot{r}_{t},\,\dot{\mathbf{x}}_{t}\,) \; > \; U_i\,(\,\ddot{r}_{t},\,\ddot{\mathbf{x}}_{t}\,), \qquad \forall \,i \in \mathbf{A}.
\end{equation}

\noindent The $i$-th agent is indifferent between the two consumption bundles if and only if:
\begin{equation}
(\,\dot{r}_{t},\,\dot{\mathbf{x}}_{t}\,) \; {\sim}_i \; (\,\ddot{r}_{t},\,\ddot{\mathbf{x}}_{t}\,)
\quad \Longleftrightarrow \quad U_i\,(\,\dot{r}_{t},\,\dot{\mathbf{x}}_{t}\,) \; = \; U_i\,(\,\ddot{r}_{t},\,\ddot{\mathbf{x}}_{t}\,), \qquad \forall \,i \in \mathbf{A}.
\end{equation}
\end{lemma}

\subsection{Market with only fundamental traders}
\begin{proposition}[strategy of fundamental traders]  \label{fund:strategy}
A seller $i\in\mathbf{FS}$ with endowment $(\,\hat{r}_{t}^{i},\,\hat{\mathbf{x}}_{t}^{i}\,)$ will be willing at date $t$ to receive for a given vector of future contingent consumption $\mathbf{x}_{t}$ no less than the reservation price $\text{\textit{Ask}}_i\,(\mathbf{x}_{t})$, such that:
\begin{gather} U_i\,\big(\,\hat{r}_{t}^{i} +\text{\textit{Ask}}_i\,(\mathbf{x}_{t}),\;\hat{\mathbf{x}}_{t}^{i} -\mathbf{x}_{t}\,\big) \;\, =\;\, U_i\,(\,\hat{r}_{t}^{i},\,\hat{\mathbf{x}}_{t}^{i}\,), \notag \\
\label{seller} E [\mathbf{x}_{t}] > 0 ,\qquad \text{\textit{Ask}}_i (\mathbf{x}_{t}) > 0 ,\qquad \forall \,i \in \mathbf{FS}.
\end{gather}

A buyer $j\in\mathbf{FB}$ with endowment $(\,\hat{r}_{t}^{j},\,\hat{\mathbf{x}}_{t}^{j}\,)$ will be willing at date $t$ to pay for a given vector of future contingent consumption $\mathbf{x}_{t}$ no more than the reservation price $\text{\textit{Bid}}_j\,(\mathbf{x}_{t})$, such that:
\begin{gather} 
U_j\,\big(\,\hat{r}_{t}^{j} -\text{\textit{Bid}}_j\,(\mathbf{x}_{t}),\;\hat{\mathbf{x}}_{t}^{j} +\mathbf{x}_{t}\,\big) \;\,=\;\, U_j\,(\,\hat{r}_{t}^{j},\,\hat{\mathbf{x}}_{t}^{j}\,), \notag \\
\label{buyer} E [\mathbf{x}_{t}] > 0 ,\qquad \text{\textit{Bid}}_j (\mathbf{x}_{t}) > 0 ,\qquad \forall \,j \in \mathbf{FB}.\end{gather}
\end{proposition}

\begin{proof}
According to the rationality axioms, preferences are strictly increasing:
\begin{equation}
(\,\hat{r}_{t}^{i},\,\hat{\mathbf{x}}_{t}^{i}\,) \; {\succ}_i \; (\,\hat{r}_{t}^{i},\;\hat{\mathbf{x}}_{t}^{i}\, -\mathbf{x}_{t}\,) , \qquad \; E [\mathbf{x}_{t}] > 0 , \qquad \forall \,i \in \mathbf{A}.
\end{equation}

\noindent For the property of continuity in preference relations, no consequence is infinitely better or infinitely worse than any other. Thus, if we add a positive amount of certain current consumption $\epsilon$ to the right-hand side of the relation, there will exist one and only one value of $\epsilon$ such that the $i$-th agent will be indifferent between the two consumption bundles:
\begin{equation} \label{epsilon}
\exists \,\epsilon>0 \quad \Longrightarrow \quad (\,\hat{r}_{t}^{i},\,\hat{\mathbf{x}}_{t}^{i}\,) \; {\sim}_i \; (\,\hat{r}_{t}^{i}\,+\epsilon,\;\hat{\mathbf{x}}_{t}^{i}\, -\mathbf{x}_{t}\,) .
\end{equation}

\noindent Let us denote $\epsilon =\text{\textit{Ask}}_i\,(\mathbf{x}_{t})$. Then:
\begin{equation}
\exists \,\epsilon>0 \quad \Longrightarrow \quad (\,\hat{r}_{t}^{i},\,\hat{\mathbf{x}}_{t}^{i}\,) \; {\sim}_i \; \big(\,\hat{r}_{t}^{i}\,+\text{\textit{Ask}}_i\,(\mathbf{x}_{t}),\;\hat{\mathbf{x}}_{t}^{i}\, -\mathbf{x}_{t}\,\big) , 
\end{equation}
\noindent which, according to lemma \ref{preference}, is equivalent to formula (\ref{seller}):
\begin{equation}
U_i\,(\,\hat{r}_{t}^{i},\,\hat{\mathbf{x}}_{t}^{i}\,) \; = \; U_i\,\big(\,\hat{r}_{t}^{i} +\text{\textit{Ask}}_i\,(\mathbf{x}_{t}),\;\hat{\mathbf{x}}_{t}^{i}\, -\mathbf{x}_{t}\,\big) .
\end{equation}

\noindent The proof of formula (\ref{buyer}) logically follows from that of formula (\ref{seller}). $\Box$
\end{proof}

\begin{corollary} 
\label{tradesellerbuyer}
The exchange of a vector $\mathbf{x}_{t}$ between a seller $i\in\mathbf{FS}$ and a buyer $j~\in~\mathbf{FB}$ can happen only at a price bounded between the reservation prices of the seller, $\text{\textit{Ask}}_i\,(\mathbf{x}_{t})$, and of the buyer, $\text{\textit{Bid}}_j\,(\mathbf{x}_{t})$:
\begin{equation}
\exists \, \text{\textit{Equil}}(\mathbf{x}_{t}) \, \Rightarrow \,  \text{\textit{Ask}}_i(\mathbf{x}_{t}) \,\leq\, \text{\textit{Equil}}(\mathbf{x}_{t}) \,\leq\, \text{\textit{Bid}}_j(\mathbf{x}_{t}),\qquad \forall i \in \mathbf{FS} \land \forall j \in \mathbf{FB}.
\end{equation}
\end{corollary}

\begin{proof}
A necessary condition for the exchange is that both seller and buyer prefer to their initial endowment the consumption bundle resulting from the trade:
\begin{equation}
\exists \, \text{\textit{Equil}} (\mathbf{x}_{t}) \, \Longrightarrow \,
\left\{ \begin{array}{ll}
\big(\,\hat{r}_{t}^{i}\,+\text{\textit{Equil}} (\mathbf{x}_{t}),\;\hat{\mathbf{x}}_{t}^{i}\, -\mathbf{x}_{t}\,\big) \, {\succsim}_i \, (\,\hat{r}_{t}^{i},\,\hat{\mathbf{x}}_{t}^{i}\,), & \; \forall\, i \in \mathbf{FS}\\
{}\\
\big(\,\hat{r}_{t}^{j}\,-\text{\textit{Equil}} (\mathbf{x}_{t}),\;\hat{\mathbf{x}}_{t}^{j}\, +\mathbf{x}_{t}\,\big) \, {\succsim}_j \, (\,\hat{r}_{t}^{j},\,\hat{\mathbf{x}}_{t}^{j}\,), & \; \forall\, j \in \mathbf{FB}.\\
\end{array} \right. \end{equation}

\noindent From formulae (\ref{seller}) and (\ref{buyer}), the condition is equivalent to corollary \ref{tradesellerbuyer}:
\begin{equation} \label{proofsellerbuyer1}
\exists \, \text{\textit{Equil}}\,(\mathbf{x}_{t}) \, \Longrightarrow \,
\left\{ \begin{array}{ll}
\text{\textit{Equil}}\,(\mathbf{x}_{t})\; \geq \; \text{\textit{Ask}}_i\,(\mathbf{x}_{t}), & \; \forall\, i \in \mathbf{FS}\\
{}\\
\text{\textit{Equil}}\,(\mathbf{x}_{t})\; \leq \; \text{\textit{Bid}}_j\,(\mathbf{x}_{t}), & \; \forall\, j \in \mathbf{FB}. \; \Box\\
\end{array} \right. 
\end{equation} \end{proof}

An almost trivial consequence of the previous propositions is that, in a market with only fundamental traders, 
 the price process of any tradable vector $\mathbf{x}_{t} \in \mathbf{x}_{t}^{set}$ is bounded. 
Its market-clearing price, $\text{\textit{Equil}}(\mathbf{x}_{t})$, assumes values in the closed interval between the minimum 
reservation price among the sellers and the maximum reservation price among the buyers, and exists if and only such interval exists and 
is positive:
\begin{gather}
\exists \, \text{\textit{Equil}}(\mathbf{x}_{t}) \;\, \Longleftrightarrow \;\,  \min[\text{\textit{Ask}}_{\mathrm{FS}}(\mathbf{x}_{t})] \;\leq\; \text{\textit{Equil}}(\mathbf{x}_{t}) \;\leq\; \max[\text{\textit{Bid}}_{\mathrm{FB}}(\mathbf{x}_{t})], \notag \\
\label{ask:bid:sets} \max[\text{\textit{Bid}}_{\mathrm{FB}}(\mathbf{x}_{t})] \;\,\geq\;\, \min[\text{\textit{Ask}}_{\mathrm{FS}}(\mathbf{x}_{t})].
\end{gather}

\begin{definition} We define the space of Pareto-efficient bargains, $\mathbf{E}_{t}$, as the set:
\begin{equation} \mathbf{E}_{t}:= \;\, \Big\{ \;\, (r_t, \,\mathbf{x}_{t}) \;\,\in\;\, \mathclose{\Big[} \, \min\,[\text{\textit{Ask}}_{\mathrm{FS}}(\mathbf{x}_{t})],\;\, \max\,[\text{\textit{Bid}}_{\mathrm{FB}}(\mathbf{x}_{t})] \, \mathclose{\Big]} \, \times\, \mathbf{x}_{t}^{set} \;\,\Big\}. \end{equation}
Any element within the space of Pareto-efficient bargains $\mathbf{E}_{t}$ denote a price $r_t$ for the asset $\mathbf{x}_{t}$, at which at least one seller and one buyer (both fundamental traders) may find convenient to make the exchange.
The set $\mathbf{E}_{t}$ can also be considered as the (closed) codomain of the function $\text{\textit{Equil}}\,(\mathbf{x}_{t}^{set})$.
\end{definition}

Let us define an arbitrage opportunity as a trading strategy that does not require the investment of current consumption, has a positive probability of gaining additional consumption and cannot lead to consumption losses. Then, the following proposition holds.

\begin{proposition} \label{primal:market} In a market with only fundamental traders, there are no arbitrage opportunities.
\end{proposition}
\begin{proof}
By assumption, the risk free interest rate in the market is zero. Then, in order to have an arbitrage opportunity, there must exists: a security that has more than one price in the market, or some replicating portfolios $\phi_t$ and $\phi'_t$ with identical cash flows and different market-clearing prices. Let us demonstrate \textit{per absurdum} that this is not possible.
Let us assume that:
\begin{equation} \label{wrong:assumption}
\exists \;\mathbf{x}_t,\, \mathbf{x}'_t \, \in \, \mathbf{x}_t^{set} \;:\;\; \mathbf{x}_t \,=\, \mathbf{x}'_t \;\; \wedge \;\; \text{\textit{Equil}}({\mathbf{x}_t}) \,\neq\, \text{\textit{Equil}}({\mathbf{x}'_t}).
\end{equation}
Then, according to formula (\ref{complete:price:function}), it should result:
\begin{equation} f\hspace{-1pt}\colon \hspace{-1pt}\text{\textit{Ask}}_{\mathrm{FS}}(\!\mathbf{x}_{t}\!) \times \text{\textit{Bid}}_{\mathrm{FB}}(\!\mathbf{x}_{t}\!) \mapsto\! \text{\textit{Equil}}(\!\mathbf{x}_{t}\!) \,\neq\, f\hspace{-1pt}\colon \hspace{-1pt}\text{\textit{Ask}}_{\mathrm{FS}}(\!\mathbf{x}'_{t}\!) \times \text{\textit{Bid}}_{\mathrm{FB}}(\!\mathbf{x}'_{t}\!) \mapsto\! \text{\textit{Equil}}(\!\mathbf{x}'_{t}\!), \end{equation} 
which for $\text{\textit{Ask}}_{\mathrm{FS}}(\mathbf{x}_{t})\times{}\text{\textit{Bid}}_{\mathrm{FB}}(\mathbf{x}_{t}) = \text{\textit{Ask}}_{\mathrm{FS}}(\mathbf{x}'_{t})\times{}\text{\textit{Bid}}_{\mathrm{FB}}(\mathbf{x}'_{t})$ is clearly false.

The proof for replicating portfolios with identical cash flows, $\phi_t$ and $\phi'_t$, logically follows from the preceding given that, for the transitivity in preferences, it results:
\begin{equation} \label{proofsellerbuyer}
\phi_{t} \approx \phi'_{t} \, \Rightarrow \,
\left\{ \begin{array}{ll}
\text{\textit{Ask}}_i(\phi_{t}) = \text{\textit{Ask}}_i(\phi'_{t}),\; \forall\, i \in \mathbf{FS}  & \; \Leftrightarrow \text{\textit{Ask}}_{\mathrm{FS}}(\phi_{t}) = \text{\textit{Ask}}_{\mathrm{FS}}(\phi'_{t}) \\
{}\\
\text{\textit{Bid}}_j(\phi_{t}) = \text{\textit{Bid}}_j(\phi'_{t}),\; \forall\, j \in \mathbf{FB}  & \; \Leftrightarrow \text{\textit{Bid}}_{\mathrm{FB}}(\phi_{t}) = \text{\textit{Bid}}_{\mathrm{FB}}(\phi'_{t}). \; \Box\\
\end{array} \right. 
\end{equation} \end{proof}

\subsection{Market with fundamental and technical traders}
In this section we demonstrate that, in a market with fundamental traders, there exists a martingale equivalent probability measure that permits the risk neutral valuation of all securities in the market. 
The possibility of risk neutral valuation leads technical traders to enter in the market.
For improving the clarity of the demonstration, we assume in the following that the Brownian motion is \mbox{$1$-dimensional}. 

\begin{definition}[technical traders' assumption] \label{tech:assumption}
Let $B_{t}$ be a $1$-dimensional Brownian motion with respect to $\mathcal{P}$, and $H_s$ an adapted c\`adl\`ag process. The price process 
$\text{\textit{Equil}} (\mathbf{x}_t)$ can be approximated with the semimartingale $Z_t$ on $(\Omega, \mathcal{F}_t, \mathcal{P})$, 
such that:
\begin{gather} \label{brownian:motion} \text{\textit{Equil}} (\mathbf{x}_t) \,\approx\, Z_t \,=\, B_t + \int_0^t H_s\,ds, \\
\text{with: } B_t \in \mathcal{M}_{loc}^c (\mathcal{P}) := \{\text{all continuous local martingales with respect to $\mathcal{P}$}\}, \notag \\
\text{and: } \int_0^t H_s\,ds \,\in\, FV \,:=\, \{\text{all stochastic processes with finite variation}\}. \mbox{\hspace{2cm}} \notag
\end{gather} 
\end{definition}

\noindent By adopting this approximation, agents can evaluate the security $\mathbf{x}_t$ via the sophisticated methods of martingale theory and stochastic calculus.

\begin{proposition}[existence of an equivalent martingale measure]\label{main:proposition}
Let the price process $Z_t$ be a semimartingale of $1$-dimensional Brownian motion under the measure~$\mathcal{P}$. 
Then, there exists a probability measure $\mathcal{Q}$, equivalent 
 to $\mathcal{P}$, such that $\text{\textit{Equil}} (\mathbf{x}_t) \approx Z_t$ can be represented as a continuous local martingale 
under $\mathcal{Q}$.
\begin{equation}
\exists \, \mathcal{Q} \sim \mathcal{P}\;: \;\;\text{\textit{Equil}} (\mathbf{x}_t) \;\approx\; Z_t \;=\; B_t + \int_0^t H_s\,ds 
\;\; \in \;\mathcal{M}_{loc}^{c}(\mathcal{Q}),\end{equation}
with $B_t\in\mathcal{M}_{loc}^c(\mathcal{P})$ \,and\, $\int_0^t H_s\,ds \in FV$.
\end{proposition}

\begin{proof} The proposition is equivalent to the Girsanov theorem, that we briefly demonstrate. 
Let be $\mathcal{Q}=\mathcal{E}(L)_t\cdot\mathcal{P}$, and let $\mathcal{E}(L)_t$ be the Radon-Nikodym derivative defined as:
\begin{equation} \mathcal{E}(L)_t = \exp \Big\{ L_t - \frac{1}{2} \left\langle L \right\rangle_t \Big\}, \end{equation}
where $\left\langle L \right\rangle_t$ is the quadratic variation of $L_t$ and:
\begin{equation} L_t = -\int_0^t H_s\,dB_s. \end{equation}
It results $E_{\mathcal{P}}[\mathcal{E}(L)_t]\equiv 1$. Moreover, $\mathcal{E}(L)_t$ is a martingale, because it fulfills the sufficient Novikov condition:
\begin{equation} E \Big[ \exp \Big\{ \frac{1}{2} \left\langle L \right\rangle_t \Big\} \Big] < \infty, \qquad \forall\,t. \end{equation}
According to the Girsanov theorem, it results:
\begin{equation} \label{eqn:girsanov}  B_t^{\mathcal{Q}} = B_t - \left\langle B, L \right\rangle_t \in \mathcal{M}_{loc}^c (\mathcal{Q}), \end{equation}
where $\left\langle B, L \right\rangle_t$ is the covariation of $B_t$ and $L_t$, with:
\begin{equation} - \left\langle B, L \right\rangle_t = \Big\langle B,\,\int_0^t H_s\,dB_s  \Big\rangle_t =
\int_0^t H_s\,d\left\langle B \right\rangle_s = \int_0^t H_s\,ds.
\end{equation}
Then, it results:
\begin{equation} - \left\langle B, L \right\rangle_t =  \int_0^t H_s\,ds \quad \Longleftrightarrow \quad B_t^{\mathcal{Q}} = Z_t \in \mathcal{M}_{loc}^c (\mathcal{Q}). 
\;\Box
\end{equation} \end{proof}

From the preceding proof, it follows that $(Z_t)_{t\in[0,T]}$ is square integrable and that, according to L\'evy theorem, 
it is a Brownian motion and its paths are of unbounded total variation with respect to $\mathcal{Q}$.

Please note that, in a market with only fundamental traders, the absence of arbitrage and free lunch opportunities 
can be considered equivalent to the existence of an equivalent probability measure $\mathcal{Q}$ under which the price processes 
of all securities become a martingale.\footnote{For a proof in markets with similar topological conditions, see for example: Kreps \cite{kre}, Harrison and Pliska \cite{har:pli}, Heath and Jarrow \cite{hea:jar}, and Delbaen \cite{del}.}

According to propositions \ref{primal:market} and \ref{main:proposition}, the existence of a market-clearing price in a market with fundamental traders leads to the existence of a risk neutral probability measure $Q_t$. This permits the entrance of technical traders in the market. 

\begin{definition}[strategy of technical traders] \label{technical trader's strategy} 
We have given a security~$\mathbf{x}_t$ whose price process $\text{\textit{Equil}}(\mathbf{x}_t)$ can be approximated with a semimartingale~$Z_t$.
Then, risk neutral agents 
will be indifferent whether to trade or not the security: 
\begin{gather}
\text{\textit{Equil}}(\mathbf{x}_t) \;\,\approx\;\, Z_t \quad\Longrightarrow\quad E_{\mathcal{P}_t} \Big[ \text{\textit{Equil}} (\mathbf{x}_{t+dt}) \Big] \;\,=\;\, E_{\mathcal{Q}} \Big[ Z_{t+dt} \Big] \quad\Longrightarrow \notag \\
\Longrightarrow\ \,
\Big( \, \hat{r}_{t}^{i} \,-\text{\textit{Equil}}(\mathbf{x}_t)\hspace{-1pt},\; \hat{\mathbf{x}}_{t}^{i} \,+\mathbf{x}_t \, \Big) \;\,{\sim}_i\;\hspace{1pt}
\Big( \, \hat{r}_{t}^{i} \,+\text{\textit{Equil}}(\mathbf{x}_t)\hspace{-1pt},\; \hat{\mathbf{x}}_{t}^{i} \,-\mathbf{x}_t \, \Big) \;\,{\sim}_i\;\hspace{1pt}
\Big( \, \hat{r}_{t}^{i},\; \hat{\mathbf{x}}_{t}^{i} \, \Big)\hspace{-1pt}, \notag \\
\forall \,i \in \mathbf{TB} \cup \mathbf{TS}.
\end{gather}

\noindent Risk neutral agents who have chosen to buy the security $\mathbf{x}_t$ will be willing to pay a price \mbox{$\text{\textit{Buy}}(\mathbf{x}_t)>\text{\textit{Equil}}(\mathbf{x}_t)$}:
\begin{gather} 
\text{\textit{E\hspace{-1pt}quil}}(\hspace{-1pt}\mathbf{x}_t\hspace{-1pt}) \hspace{1pt}\approx\hspace{1pt} Z_t \;\,\Rightarrow\;\, E_{\mathcal{\hspace{-1pt}Q}} \Big[ Z_{t+dt} \Big] \hspace{1pt}<\hspace{1pt} E_{\mathcal{P}_t} \hspace{-1pt}\Big[ \text{\textit{E\hspace{-1pt}quil}} (\hspace{-1pt}\mathbf{x}_{t+dt}\hspace{-1pt}) \Big| \begin{smallmatrix} \text{decision of} \\ \text{buying $\mathbf{x}_t$} \end{smallmatrix} \Big] \hspace{1pt}=\hspace{1pt} \text{\textit{B\hspace{-1pt}uy}}(\hspace{-1pt}\mathbf{x}_t\hspace{-1pt}) \;\,\Rightarrow \notag \\
\Longrightarrow\ \,
\Big(\hspace{1pt} \hat{r}_{t}^{i} \,-\text{\textit{B\hspace{-1pt}uy}}(\hspace{-1pt}\mathbf{x}_t\hspace{-1pt})\hspace{-1pt},\,\hspace{1pt} \hat{\mathbf{x}}_{t}^{i} \,+\mathbf{x}_t \hspace{1pt}\Big) \;\,{\sim}_i\;
\Big(\hspace{1pt} \hat{r}_{t}^{i} \,-\text{\textit{E\hspace{-1pt}quil}}(\hspace{-1pt}\mathbf{x}_t\hspace{-1pt}) -\epsilon,\,\hspace{1pt} \hat{\mathbf{x}}_{t}^{i} \,+\mathbf{x}_t \hspace{1pt} \Big) \;\,{\succ}_i\;
\Big(\hspace{1pt} \hat{r}_{t}^{i},\,\hspace{1pt} \hat{\mathbf{x}}_{t}^{i} \hspace{1pt}\Big)\hspace{-1pt}, \notag \\
\forall \,i \in \mathbf{TB}, \; \epsilon > 0. \label{strategy:buy}
\end{gather}

\noindent Risk neutral agents who have chosen to sell the security $\mathbf{x}_t$ will be willing to accept a price \mbox{$\text{\textit{Sell}}(\mathbf{x}_t)<\text{\textit{Equil}}(\mathbf{x}_t)$}:
\begin{gather} 
\text{\textit{E\hspace{-1pt}quil}}(\hspace{-1pt}\mathbf{x}_t\hspace{-1pt}) \hspace{1pt}\approx\hspace{1pt} Z_t \;\,\Rightarrow\;\, E_{\mathcal{\hspace{-1pt}Q}} \Big[ Z_{t+dt} \Big] \hspace{1pt}>\hspace{1pt} E_{\mathcal{P}_t} \hspace{-1pt}\Big[ \text{\textit{E\hspace{-1pt}quil}} (\hspace{-1pt}\mathbf{x}_{t+dt}\hspace{-1pt}) \Big| \begin{smallmatrix} \text{decision of} \\ \text{selling $\mathbf{x}_t$} \end{smallmatrix} \Big] \hspace{1pt}=\hspace{1pt} \text{\textit{S\hspace{-1pt}ell}}(\hspace{-1pt}\mathbf{x}_t\hspace{-1pt}) \;\,\Rightarrow \notag \\
\Longrightarrow \,
\Big(\hspace{1pt} \hat{r}_{t}^{j} \,+\text{\textit{S\hspace{-1pt}ell}}(\hspace{-1pt}\mathbf{x}_t\hspace{-1pt})\hspace{-1pt},\,\hspace{1pt} \hat{\mathbf{x}}_{t}^{j} \,-\mathbf{x}_t \hspace{1pt}\Big) \;\,{\sim}_j\;
\Big(\hspace{1pt} \hat{r}_{t}^{j} \,+\text{\textit{E\hspace{-1pt}quil}}(\hspace{-1pt}\mathbf{x}_t\hspace{-1pt}) +\epsilon,\,\hspace{1pt} \hat{\mathbf{x}}_{t}^{j} \,-\mathbf{x}_t \hspace{1pt} \Big) \;\,{\succ}_{\hspace{-1pt}j}\;
\Big(\hspace{1pt} \hat{r}_{t}^{j},\,\hspace{1pt} \hat{\mathbf{x}}_{t}^{j} \hspace{1pt}\Big)\hspace{-1pt}, \notag \\
 \forall \,j \in \mathbf{TS}, \; \epsilon > 0. \label{strategy:sell}
\end{gather}
\end{definition}

As exposed in section \ref{agents}, the element $\epsilon$ in formulae (\ref{strategy:buy}) and (\ref{strategy:sell}) can be interpreted as the Bayesian updating of price expectations of technical agents due to their choice of buying or selling. Moreover, we can interpret $\epsilon$ as a small sum that agents should add or subtract to $Equil\,(\mathbf{x}_{t})$ to make sure that their market orders would be executed.

Risk neutral investors who trades randomly, according to the martingale equivalent probability measure inferred from market prices, can be seen as a particular kind of noise traders. They differ from the classical definition of noise traders in literature\footnote{See for example: Black \cite{bla86} and De Long et al. \cite{del:shl:sum:wal}.} in the fact that they are rational and informed.
A consequence of proposition \ref{main:proposition} and definition \ref{technical trader's strategy} is that the market price function is no longer unbounded if technical traders are operating. This leads to the following proposition.

\begin{proposition} \label{value:outside:pareto} In a market with fundamental and technical traders, the price of the security $\mathbf{x}_t$ can assume a value outside the set of Pareto-efficient bargains of fundamental traders.
\begin{equation}
\text{\textit{Equil}}(\hspace{-1pt}\mathbf{x}_t\hspace{-1pt}) \!\approx\! Z_t \;\hspace{1pt} \land \;\hspace{1pt} \exists\hspace{1pt} i \hspace{-2pt}\in\hspace{-2pt} \mathbf{TB}_t \; \land \; \exists\hspace{1pt} j \hspace{-2pt}\in\hspace{-2pt} \mathbf{TS}_t \;\;\; \Rightarrow \;\;\;
P \big[ \text{\textit{Equil}}(\hspace{-1pt}\mathbf{x}_t\hspace{-1pt}) \notin \mathbf{E}_{t} \big] > 0,
\end{equation}
\noindent where $\mathbf{TB}_t$ and $\mathbf{TS}_t$ represent, respectively, the sets of technical buyers and technical sellers at date $t$.
\end{proposition}

\begin{proof}
Let us suppose that the clearing-market price for a security $\mathbf{x}_t$ exists at date $t$ and is equal to the highest reservation price among buyers that are fundamental traders, $\max[\text{\textit{Bid}}_{\mathrm{FB}}(\mathbf{x}_{t})]$. All sellers and at least one buyer among fundamental traders are then operating; also, according to definition \ref{technical trader's strategy}, risk neutral investors can choose to trade according to the martingale equivalent probability measure that emerges from market prices. 

Let us now suppose that at time $t$ the number of technical traders for which:
\begin{equation}
\text{\textit{Equil}} (\tilde{\mathbf{x}}_t) \approx Z_t,\;\, E_{\mathcal{P}_t} \big[ \text{\textit{Equil}} (\mathbf{x}_{t+dt}) \big| \begin{smallmatrix} \text{decision of} \\ \text{buying $\mathbf{x}_t$} \end{smallmatrix} \big] \,>\, E_{\mathcal{Q}} \big[ Z_{t+dt} \big],\;\, \forall \,i \in \mathbf{TB}_t, \end{equation}
exceeds the number of technical traders for which:
\begin{equation}
\text{\textit{Equil}} (\tilde{\mathbf{x}}_t) \approx Z_t,\;\, E_{\mathcal{P}} \big[ \text{\textit{Equil}} (\mathbf{x}_{t+dt}) \big| \begin{smallmatrix} \text{decision of} \\ \text{selling $\mathbf{x}_t$} \end{smallmatrix} \big] \,<\, E_{\mathcal{Q}} \big[ Z_{t+dt} \big],\;\, \forall \,j \in \mathbf{TS}_t. \end{equation}

\noindent This means that the number of technical traders who are willing to buy the security $\mathbf{x}_t$ at the market-clearing price, exceeds at date $t$ the number of technical traders who are willing to sell it. Thus, by assumption in section \ref{sec:orders} and in definition \ref{technical trader's strategy}, the market-clearing price raises; then it can exceed $\max[\text{\textit{Bid}}_{\mathrm{FB}}(\mathbf{x}_{t})]$ and it can exit form the space of Pareto-efficient bargains $\mathbf{E}_t$.  
At the new price, no buyer that is fundamental trader is operating. The price will increase until technical traders who are willing to buy exceeds those who are willing to sell, and it will decrease vice versa. $\,\Box$ \end{proof}

\subsection{Market conditions} \label{market:conditions}
According to the definition of the price process in formula (\ref{complete:price:function}) and to proposition \ref{value:outside:pareto}, we can now distinguish and quantitatively define the following different market conditions:
\begin{itemize}
\item \textit{non speculative market}: it happens when, at a given date $t$, only fundamental traders are operating. The market-clearing price of a security $\mathbf{x}_t$ belongs to the set of Pareto-efficient bargains $\mathbf{E}_t$. 
The price function becomes:
\begin{gather} f\colon \text{\textit{Bid}}_{\mathrm{FB}}(\mathbf{x}_t) \times \text{\textit{Ask}}_{\mathrm{FS}}(\mathbf{x}_t) \mapsto \text{\textit{Equil}}(\mathbf{x}_{t}) 
\in \mathbf{E}_t, \notag \\ \min[\text{\textit{Ask}}_{\mathrm{FS}}(\mathbf{x}_t)] \;\leq\; \text{\textit{Equil}}(\mathbf{x}_{t}) \;\leq\; \max[\text{\textit{Bid}}_{\mathrm{FB}}(\mathbf{x}_t)].
\end{gather}
\item \textit{normal market}: it happens when, at a given date $t$, both fundamental and technical traders are operating in the market and the market-clearing price of a security $\mathbf{x}_t$ belongs to the set of Pareto-efficient bargains $\mathbf{E}_t$. 
The price function results:
\begin{gather}
f\hspace{-1pt}\colon \text{\textit{Bid}}_{\hspace{-1pt}\mathrm{F\hspace{-1pt}B}}(\!\mathbf{x}_t\!) \!\times\! 
\text{\textit{A\hspace{-1pt}sk}}_{\mathrm{\hspace{-1pt}F\hspace{-1pt}S}}(\!\mathbf{x}_t\!) \!\times\! 
\text{\textit{B\hspace{-1pt}uy}}_{\mathrm{\hspace{-1pt}T\hspace{-1pt}B}}(\!\mathbf{x}_t\!) \!\times\! 
\text{\textit{\hspace{-1pt}S\hspace{-1pt}ell}}_{\mathrm{T\hspace{-1pt}S}}(\!\mathbf{x}_t\!) \!\times\! 
\text{\textit{E\hspace{-1pt}quil}}(\!\mathbf{x}_{t\hspace{-1pt}-\hspace{-1pt}dt}\!) 
\mapsto\hspace{-1pt} \text{\textit{E\hspace{-1pt}quil}}(\!\mathbf{x}_t\!) \hspace{-1pt}\in\hspace{-1pt} \mathbf{E}_t, \notag \\
\min[\text{\textit{Ask}}_{\mathrm{FS}}(\mathbf{x}_t)] \;\leq\; \text{\textit{Equil}}(\mathbf{x}_{t}) \;\leq\; \max[\text{\textit{Bid}}_{\mathrm{FB}}(\mathbf{x}_t)].
\end{gather}
\item \textit{market bubble}: it happens when, at a given date $t$, the market-clearing price of a security $\mathbf{x}_t$ exceeds the highest reservation price among buyers that are fundamental traders, $\max[\text{\textit{Bid}}_{\mathrm{FB}}(\mathbf{x}_{t})]$; thus, it is outside the set of Pareto-efficient bargains $\mathbf{E}_t$. Only technical traders and sellers that are fundamental traders are operating in the market. 
The price function becomes:
\begin{gather}
f\colon \text{\textit{Ask}}_{\mathrm{FS}}(\mathbf{x}_t) \times \text{\textit{Buy}}_{\mathrm{TB}}(\mathbf{x}_t) \times \text{\textit{Sell}}_{\mathrm{TS}}(\mathbf{x}_t) \times \text{\textit{Equil}}(\mathbf{x}_{t-dt}) \mapsto \text{\textit{Equil}}(\mathbf{x}_t) \notin \mathbf{E}_t, \notag \\
 \text{\textit{Equil}}(\mathbf{x}_{t}) \;>\; \max[\text{\textit{Bid}}_{\mathrm{FB}}(\mathbf{x}_t)].
\end{gather}
\item \textit{market depression}: it happens when, at a given date $t$, the market-clearing price of a security $\mathbf{x}_t$ is below the lowest reservation price among sellers that are fundamental traders, $\min[\text{\textit{Ask}}_{\mathrm{FS}}(\mathbf{x}_{t})]$; thus, it is outside the set of Pareto-efficient bargains $\mathbf{E}_t$. Only technical traders and buyers that are fundamental traders are operating in the market. 
The price function becomes:
\begin{gather}
f\colon \text{\textit{Bid}}_{\mathrm{FB}}(\mathbf{x}_t) \times \text{\textit{Buy}}_{\mathrm{TB}}(\mathbf{x}_t) \times \text{\textit{Sell}}_{\mathrm{TS}}(\mathbf{x}_t) \times \text{\textit{Equil}}(\mathbf{x}_{t-dt}) \mapsto \text{\textit{Equil}}(\mathbf{x}_t) \notin \mathbf{E}_t, \notag\\
 \text{\textit{Equil}}(\mathbf{x}_{t}) \;<\; \min[\text{\textit{Ask}}_{\mathrm{FS}}(\mathbf{x}_t)].
\end{gather}
\end{itemize}

\subsection{The risk neutral valuation paradox} \label{main:demonstration}
\begin{proposition} \label{jump} If technical traders stop operating in a market which is under bubble conditions or under depression conditions, the market-clearing price of the security $\mathbf{x}_t$ will have a jump discontinuity. The new price  will assume a value inside the set of Pareto-efficient bargains of fundamental traders.
\end{proposition}

\begin{proof}
We first prove the proposition under market bubble conditions.
Let us suppose that, at date $t$, the market-clearing price of the security $\mathbf{x}_{t}$ is outside the space of Pareto-efficient bargains $\mathbf{E}_t$ and exceeds the highest reservation price among buyers that are fundamental traders, $\max[\text{\textit{Bid}}_{\mathrm{FB}}(\mathbf{x}_{t})]$. Thus, the market is experiencing a bubble.

Let us suppose that, while the market is still under bubble conditions, at some date $\tilde{t}>t$ all technical traders stop operating. 
This event has a positive probabilities because the number of technical traders operating in the market varies randomly over time.
The left and right limits of the price function in the neighborhood of $\tilde{t}$ result:
\begin{gather}
\lim_{t\rightarrow \tilde{t}^{-}} \Big[ f\colon \!\text{\textit{Ask}}_{\mathrm{FS}}(\mathbf{x}_{\tilde{t}^{-}}) \!\times\! \text{\textit{Buy}}_{\mathrm{TB}}(\mathbf{x}_{\tilde{t}^{-}}) \!\times\! \text{\textit{Sell}}_{\mathrm{TS}}(\mathbf{x}_{\tilde{t}^{-}}) \!\times\! \text{\textit{Equil}}(\mathbf{x}_{t-dt}) \mapsto \text{\textit{Equil}}(\mathbf{x}_{\tilde{t}^{-}}) \Big], \notag \\
\label{eqn:bubble}
\mbox{\hspace{0cm}} \text{\textit{Equil}}(\mathbf{x}_{\tilde{t}^{-}}) \;>\; \max[\text{\textit{Bid}}_{\mathrm{FB}}(\mathbf{x}_{\tilde{t}^{-}})],
\end{gather}
\noindent and:
\begin{gather}
\lim_{t\rightarrow \tilde{t}^{+}} \Big[ f\colon \text{\textit{Bid}}_{\mathrm{FB}}(\mathbf{x}_{\tilde{t}^{+}}) \times \text{\textit{Ask}}_{\mathrm{FS}}(\mathbf{x}_{\tilde{t}^{+}}) \mapsto \text{\textit{Equil}}(\mathbf{x}_{\tilde{t}^{+}}) \Big], \notag \\
 \min[\text{\textit{Ask}}_{\mathrm{FS}}(\mathbf{x}_{\tilde{t}^{+}})] \;\leq\; \text{\textit{Equil}}(\mathbf{x}_{\tilde{t}^{+}}) \;\leq\; \max[\text{\textit{Bid}}_{\mathrm{FB}}(\mathbf{x}_{\tilde{t}^{+}})].
\end{gather}
\noindent Then, the price function has a jump discontinuity at date $\tilde{t}$. Indeed, it results:
\begin{equation}
\text{\textit{Equil}}(\mathbf{x}_{\tilde{t}^{-}}) \,\neq\, \text{\textit{Equil}}(\mathbf{x}_{\tilde{t}^{+}}). \end{equation}
\noindent The jump consists in a negative price shock:
\begin{equation}
\text{\textit{Equil}}(\mathbf{x}_{\tilde{t}^{-}}) \;>\; \text{\textit{Equil}}(\mathbf{x}_{\tilde{t}^{+}}).
\end{equation}

\noindent The proposition is then proved under market bubble conditions. It can be proved analogously under market depression. $\;\Box$ \end{proof}

\noindent Please note that the market-clearing price $\text{\textit{Equil}}(\mathbf{x}_t)$ will have a jump discontinuity also if all technical traders who are, alternatively, buyers or sellers will stop operating in a market which is, respectively, under bubble or depression conditions.


\begin{proposition}[risk neutral valuation paradox] Let us consider an incomplete market for derivatives, which is frictionless, informationally efficient, and has competitive rational investors with different risk attitudes. Then, if some investors are risk neutral, the market may exhibit almost surely cycles in which prices diverge from their fundamental values, ended by abrupt price adjustments towards their fundamental values.
\end{proposition}

\begin{proof}
The demonstration logically follows from proposition \ref{jump}. $\;\Box$ \end{proof}

\section{Conclusions} \label{conclusions}
In this paper we have highlighted the role of risk neutral investors in generating endogenous bubbles in derivatives markets.
In an incomplete market for derivatives which has no arbitrage opportunities, it is indifferent to risk neutral investors whether to trade or not and, 
in the first case, whether to buy or to sell any particular security. Hence, their trading choices will vary randomly over time. 
Then, prices can diverge from their fundamental values but, whenever risk neutral investors stop trading, prices may have abrupt adjustments towards their fundamental values.

Our main result consists in the demonstration of the existence of a price anomaly, that we have called the risk neutral valuation paradox.
The paradox is a consequence of the fundamental theorem of asset pricing and the related martingale theory of bubbles. 
It states that derivatives markets, which are informationally efficient, incomplete and with some risk neutral investors, may exhibit violations to the Gaussian random walk hypothesis.
Furthermore, it suggests that extreme price movements like price peaks or crashes may happen with a higher than-normal frequency.

The paradox may yield surprisingly concrete implications for the theory of financial and derivatives markets. We suggest four possible advances. The first is that violations to the random walk hypothesis can be compatible with informationally efficient markets.
The second is that the informative content of prices in almost perfect markets may decrease when risk neutral investors are operating.
The third is that risk neutral investors, even when they are rational and informed, may stimulate the emergence of boom-and-bust cycles. The fourth is that, under some circumstances, the distribution of price movements in an almost perfect market may differ from a normal distribution.

A number of caveats need to be noted regarding the present study. 
The most important limitation lies in the fact that we have demonstrated the existence of the paradox in a rather general setting. 
Indeed, a more precise definition of the price process will be necessary to extend the results.
Further work needs to be done to specify more carefully how the random variations of the prices would look over time 
and to establish to what extent the paradoxical effect is significant.

\end{document}